\documentstyle[aps,prl,floats,epsfig]{revtex}
\draft

\def\beq{\begin{equation}}
\def\eeq{\end{equation}}
\def\eea{\end{eqnarray}}
\def\bea{\begin{eqnarray}}

\def\Nc{N_c}
\def\Agroup{A}
\def\Bgroup{B}
\def\Nf{N_f}
\def\ov{/}
\begin{document}

\twocolumn[
\hsize\textwidth\columnwidth\hsize\csname @twocolumnfalse\endcsname

\title{Constraints on ultraviolet-stable fixed points in supersymmetric 
gauge theories}
\author{Stephen P. Martin$^a$ and James D. Wells$^b$}
\address{$^{(a)}$Department of Physics, Northern Illinois University, 
DeKalb~IL~60115  {\rm and} \\  
Fermi National Accelerator Laboratory, PO Box 500, Batavia~IL~60510 \\
$^{(b)}$Physics Department, University of California, Davis~CA~95616,
{\rm and}\\ 
Theory Group, Lawrence Berkeley National Laboratory, Berkeley~CA~94720}
\maketitle

\begin{abstract}  We consider the possibility that some supersymmetric
gauge theories which are not asymptotically free can be governed by
ultraviolet-stable fixed points. If this scenario can be realized, gaugino
masses will exhibit power-law running with scale, providing a possible
solution to the supersymmetric flavor problem. While naive perturbative
calculations hint at the appearance of ultraviolet-stable fixed points in
certain theories, there are strong constraints following from limits on
the scaling dimensions of gauge-invariant operators, positivity of central
charges, and Cardy's conjectured constraint on the flow of the Euler
coefficient in the stress tensor trace anomaly. Also, we prove that if
ultraviolet-stable fixed points do exist, they cannot occur in the
perturbative regime of a renormalizable model, and that all-orders results
in the limit of large numbers of chiral superfields are necessarily
inconclusive. However, we argue that the general idea of
ultraviolet-stable fixed points in supersymmetric gauge theories is
viable, and exhibit models that can satisfy all known constraints.   
\end{abstract}  

\vspace{-0.18in}
  
\pacs{hep-ph/0011382, FERMILAB-Pub-00/301-T, LBNL-47122}

\vspace{0.18in}

]
\setcounter{footnote}{1}


\section{Introduction}

Our understanding of the high-energy behavior of gauge couplings in
supersymmetric (SUSY) theories depends crucially on the matter content of the
model. If the number of chiral superfields is sufficiently small,  the
one-loop beta function for gauge couplings is negative, and the
ultraviolet (UV) theory is formally governed by a free fixed point in
which the renormalizable gauge couplings are driven to zero. 

In constrast, it is usually assumed that  supersymmetric gauge theories
with positive one-loop $\beta$ functions have a Landau pole in the UV. The
scale of the Landau pole may represent the energy above which new physics
must enter in the form of e.g.~extra dimensions, string interactions, or
quantum gravity. Above that scale, either the running of the gauge
coupling is modified so that it remains finite, or else the gauge coupling
has no simple meaning within the new framework in the far UV. In either
case, it is quite difficult to  understand in detail how to match the
model to a more fundamental theory.  There are many otherwise-attractive
extensions of the minimal supersymmetric standard models (MSSM)
\cite{primer} that seem to fit into this category. For example, the
one-loop beta functions are large and positive in many supersymmetric
grand unified theories (GUTs) and in string-inspired models with large
numbers of vectorlike representations.

In this paper, we will consider another possibility,  the idea that a
renormalizable supersymmetric gauge theory may have a non-trivial
UV-stable fixed point associated with a theory with  exact or approximate
superconformal invariance. In this scenario, the UV theory approaches a
critical point in which (at least some) supersymmetric gauge and Yukawa
couplings are nearly stationary, and  soft supersymmetry-breaking
couplings have a power-law running with scale. The criticality may be only
approximate, if some couplings are small. What differentiates this from
the situation in the previous paragraph is that the theory can remain near
the critical point for a significant range of energy scales up to the
Planck scale (or another scale of fundamental physics such as a
compactification scale or string scale). 

If a UV fixed point can indeed be realized, it has  interesting and
unique implications. For example, we will show below that gaugino  masses
necessarily scale to zero by a power law in the UV. This enables large
hierarchies in soft masses  to be generated from renormalization group
(RG) running. These hierarchies could be used to suppress flavor-violating
interactions in supersymmetry. Another intriguing feature of the UV fixed
point idea is that it could be related to the apparent need for
semi-perturbative unification in string theory
\cite{dilaton,semiperturbative}. In string theory with weak coupling, the
dilaton potential has no stable vacuum. Strongly-coupled string theories
naively have the same problem since they are dual to weakly-coupled
theories.  Perhaps an intermediate-strength coupling of a settled and
non-trivial ($g\neq 0$) UV fixed point could be part of the solution to
this problem.

However, it is not clear whether renormalizable supersymmetric models with
UV-stable fixed points actually exist in four dimensions. 
As we will show, there are some
hints from finite-order calculations that weakly suggest such
UV-stable fixed points can occur. However, there are also strong 
constraints that follow from the general properties of superconformal
theories, from the fact that scaling dimensions of gauge-invariant operators
cannot be less than 1, from the positivity of central charges, and from the
irreversibility of renormalization group RG flows as embodied in
Cardy's
conjecture \cite{Cardy} generalizing to 4 dimensions the Zamolodchikov
$c$-theorem in 2 dimensions \cite{Zamolodchikov}. We will examine these
restrictions, and show that there are some simple models that can satisfy
all known constraints including Cardy's conjecture.

\section{The NSVZ $\beta$ function, anomalous dimensions, and 
$R$-charges at superconformal fixed points}

We begin by discussing some general properties that must
be satisfied by fixed points in a supersymmetric gauge theory.
The existence of a fixed point should be 
independent of the renormalization scheme used.
However, it is most convenient to use the NSVZ scheme, since there is an
exact relation between beta functions and
the anomalous dimensions and $R$-charges of chiral superfields.
In our notation, the NSVZ gauge coupling $\beta$ function is given by
\beq
\beta(g) = {g^3 \over 16\pi^2} \left [ - 3 C_G 
+ \sum_r I_r (1 - 2 \gamma_r) 
\over
1 - C_G g^2/8 \pi^2\right ] \, ,
\label{betaNSVZ}
\eeq
where $C_G$ is the Dynkin index of the adjoint representation, the sum
$\sum_r$ is over the irreducible representations of the chiral superfields
with chiral superfield anomalous dimensions $\gamma_r$,
and $I_r$ is the Dynkin index of the representation $r$. The normalizations 
are such that $C_G = \Nc$ and $I_r = 1$ for each fundamental
plus anti-fundamental
representation of $SU(\Nc)$. The $\beta$
function has a pole at very strong coupling: $g^2_{\rm pole} = 8
\pi^2/C_G$, but we assume that this plays no role and
any fixed points occur at weaker coupling $g_* < g_{\rm pole}$. 
The numerator of $\beta(g)$ must therefore vanish at a
fixed point $g=g_*$ corresponding to a scale-invariant theory.  

Although there is no proof for four dimensions,  one expects  the
equivalence of scale invariant theories and conformal 
theories~\cite{Polchinski:1988dy}. The conformal symmetries are
necessarily unified with supersymmetry transformations to form the
superconformal algebra. This implies stringent constraints, since the
superconformal symmetry includes a non-anomalous  $U(1)$ $R$-symmetry and
an invariance under scaling. 
The $R$-charge  and superconformal scaling dimension $D$  
for any gauge-invariant chiral superfield must be related by
\cite{Dobrev:1985qv}
\beq
D = 3R/2 \> .
\label{dconsuper}
\eeq
Since the scaling dimension for any
composite chiral superfield is the sum of $1+\gamma_r$ for its
constituents, it follows that
\beq
3(R_r -1) = 2 \gamma_{r }(g_*) -1 
\label{Rgamma}
\eeq
at a superconformal fixed point $g=g_*$.  This establishes a connection
between the conditions for the vanishing of the NSVZ $\beta$ function and
the cancellation of the $R$-symmetry anomaly. In particular,  since the
$R$-charge of the chiral fermion in the representation labelled by $r$ is
$R_r -1$, and that of the gaugino is 1, the anomaly cancellation
requirement
\beq
C_G + \sum_r I_r (R_r - 1) = 0
\label{anomalycan}
\eeq
is equivalent to the vanishing of the numerator in eq.~(\ref{betaNSVZ}).
In some theories, where either symmetries or the presence of a
superpotential are sufficient to fix the $R$-charges uniquely, this allows
an exact non-perturbative determination of the anomalous dimensions at the
fixed point. More generally, it gives one linear relation between the
anomalous dimensions, if the assignment  of $R$-charges is not uniquely
determined.  Therefore, eq.~(\ref{dconsuper})  is essentially tautological
(provided that the appropriate $R$-symmetry can be uniquely determined)
rather than a non-trivial constraint in superconformal theories.

If $\beta(g_*) = 0$ and  $\beta(g) < 0$ for $0 < g < g_*$, then $g_*$
can be an IR-stable fixed point.  For example, in supersymmetric QCD with
$\Nc$ colors and $\Nf$ flavors of quark and antiquark superfields, it has
been
argued \cite{Seiberg:1995pq,Kogan:1996mr} 
that there is an IR-stable fixed point
for $3\Nc/2 < \Nf < 3\Nc$. The $R$-charges of the quark and antiquark
superfields
are both equal to $R_Q = R_{\overline Q} = 1 - \Nc/\Nf$ corresponding to
an anomalous dimension (in our normalization) $\gamma_Q(g_*) =
\gamma_{\overline Q}(g_*) = (1 - 3 \Nc/\Nf)/2$.  While the existence of
the IR-stable fixed points in the theories is conjectured, the relations
can be checked in perturbation theory when $3 \Nc - \Nf$ is very small, as
in the case of non-supersymmetric QCD \cite{Banks:1982nn}. A conjectured
dual description can be used to check the relations when $\Nf - 3 \Nc/2$
is small and positive.  
This lends credence to the picture of a pair of interacting
superconformal theories having the same physics in the IR. Other examples
include deformations of $N=2$  and finite $N=1$ theories, which generally
can have IR-stable fixed lines \cite{Leigh:1995ep}.

If instead $\beta(g_*) = 0$ and $\beta(g) > 0$ for the range $0 < g < g_*$,
then $g_*$ is a UV-stable fixed point. This is the case of interest in
this paper. Note that this implies $\partial\beta/\partial g < 0$ at the
fixed point,  a fact that will be important in the following.  It also
requires that $\sum_r I_r \geq 3 C_G$ so that the leading-order $\beta$
function is positive. The question then becomes whether theories that
satisfy this can be found with  anomalous dimensions and $R$-charges
satisfying eqs.~(\ref{Rgamma}) and (\ref{anomalycan}). Before addressing
the specific constraints on this scenario, it is convenient to first
consider what we can say about the RG running of soft
supersymmetry-breaking   parameters in a realistic version of such a
model.

\section{Evolution of soft supersymmetry-breaking parameters near a
UV-stable fixed point} 

In this section, we assume for the moment that the exact $\beta$
functions of the theory do admit a UV-stable fixed point.   
In a realistic model, the superconformal
invariance will not be exact. Let us assume that 
SUSY breaking is communicated, for example by supergravity mediation, 
to the MSSM sector at a high
scale $Q_0$ within the regime of the fixed point. In general, the soft
supersymmetry breaking mass parameters will undergo power-law RG evolution,
so that in the limit $Q \ll Q_0$,
\beq
m_{\rm soft}(Q) \sim (Q_0/Q)^k
\eeq
where $k$ is a critical exponent for the particular soft mass, and $Q$
is the RG scale. If $k>0$,
then the corresponding soft term grows as we move towards the IR within
the fixed point region. In general, we would like to investigate the
properties of the critical scaling for soft terms in realistic models.

For simplicity, we begin with the case of a model with a simple gauge group
and superpotential couplings 
sufficiently small that they can be
neglected. 
Remarkably, the exact gaugino mass beta
function is related~\cite{gauginomass} to the exact gauge coupling beta
function in the NSVZ scheme by the simple expression
\bea
\beta_M=2Mg^2 \frac{\partial}{\partial g^2} 
( \beta/g ).
\eea
At a UV-stable fixed point, $\beta=0$ and $\partial\beta/\partial g < 0$
must hold. It follows that
\bea
{dM \over dt} = \beta_M=  -\kappa M,
\label{gauginobeta}
\eea
where $t= {\rm ln}Q$ and
\beq
\kappa \equiv -\partial \beta/\partial g
\eeq 
is a positive constant in the fixed point regime, 
defined as $g$ for which $|\partial\beta/\partial g | \gg |\beta /g |$.  
Note that $\kappa$ is independent of the choice of renormalization scheme.

The scale dependence of the gaugino mass is 
\bea
M(Q)= (Q_0/Q)^\kappa M_0.
\label{gauginopowerlaw}
\eea
Thus $M$ runs according to a power law in the fixed point region.  If
$Q_0$ is the boundary condition scale where supersymmetry breaking is
transmitted, then it is possible to have $M_0$ at this high scale be very
small, yet induce large $M$ at lower scales $Q$ due to the power law
running in the fixed point region.  The gravitino mass, on the other hand,
is presumably correlated with the boundary condition gaugino mass,
$m_{3/2} \sim M_0$. Therefore, the physical gaugino can be significantly
more massive than the gravitino. The ratio of the masses is estimated as
\bea
M/m_{3/2} \sim \left( Q_0/Q_L \right)^\kappa ,
\eea
where $Q_L$ is the lowest scale of the UV fixed point regime.

The above results can be generalized to more complicated models.
As we will see in section \ref{sec:central}, it is quite possible
that the existence of a UV-stable fixed point requires the presence
of superpotential couplings. If the ultimate gauge group
is not simple, then it is possible that several gauge couplings will
be significant at the fixed point.
Therefore, let us consider the more general possibility of a 
fixed point with several distinct 
non-zero gauge couplings $g_{\Agroup}$ and non-zero holomorphic 
Yukawa couplings $Y_{ijk}$. The superpotential is
\beq
W = {1\over 6} Y^{ijk} \Phi_i \Phi_j \Phi_k . 
\label{superpot}
\eeq 
Now consider the RG running of deviations from
the fixed point: $g_{\Agroup} = g_{\Agroup *} + \delta_{\Agroup}$; 
$Y^{ijk} = Y^{ijk}_* +
\delta^{ijk}$; $Y_{ijk} = Y_{ijk*} + \delta_{ijk}$. One obtains:
\beq
{d \over dt} 
\pmatrix{\delta_{\Agroup} \cr \delta^{ijk} \cr \delta_{ijk}}
= -C
\pmatrix{ \delta_{\Bgroup} \cr \delta^{lmn} \cr \delta_{lmn}}
\eeq 
where
\beq
C =
-\pmatrix{
{\partial \beta_{\Agroup} \ov \partial g_{\Bgroup} } &
{\partial \beta_{\Agroup} \ov \partial Y^{lmn}} &  
{\partial \beta_{\Agroup} \ov \partial Y_{lmn}} \cr
{\partial \beta^{ijk} \ov \partial g_{\Bgroup} } &
{\partial \beta^{ijk} \ov \partial Y^{lmn}} &  
{\partial \beta^{ijk} \ov \partial Y_{lmn}} \cr
{\partial \beta_{ijk} \ov \partial g_{\Bgroup} } &
{\partial \beta_{ijk} \ov \partial Y^{lmn}} &
{\partial \beta_{ijk} \ov \partial Y_{lmn}} } \> .
\eeq
Here we have used a notation in which $\beta_{\Agroup} $, $\beta^{ijk}$ and
$\beta_{ijk}$ denote the $\beta$ functions of $g_{\Agroup}$, 
$Y^{ijk}$, and $Y_{ijk}$
respectively. Following common practice, complex conjugation of
couplings is denoted by raising and lowering indices,
e.g.~$Y_{ijk} = (Y^{ijk})^*$.
By definition, at a 
UV-stable fixed point, all eigenvalues of $C$ must have 
non-negative real parts. For a realistic model, we assume
that the superconformal invariance is only approximate, and in particular
is violated
by soft gaugino masses $M_{\Agroup}$ and by trilinear scalar
couplings:
\beq
{\cal L} = - {1\over 6} a^{ijk} \phi_i \phi_j \phi_k + {\rm c.c.}
\eeq
The running of the soft gaugino masses $M_{\Agroup}$ and the trilinear   
couplings $a^{ijk}$ is then given exactly in terms of the
NSVZ $\beta$ functions of the supersymmetric parameters by 
\cite{gauginomass}
\beq
{d \over dt} \pmatrix{g_{\Agroup} M_{\Agroup} \cr  a^{ijk} } = - {K}    
\pmatrix{  g_{\Bgroup} M_{\Bgroup} \cr a^{lmn} } ,
\eeq
where
\beq
{K} = \pmatrix{
-{\partial \beta_{\Agroup} \ov \partial g_{\Bgroup}} &
2{\partial \beta_{\Agroup} \ov \partial Y^{lmn}}  \cr
{\partial \beta^{ijk} \ov \partial g_{\Bgroup} } &
-2 {\partial \beta^{ijk} \ov \partial Y^{lmn}}
} \> .  
\label{Kmatrix}  
\eeq
Here we have dropped terms which vanish at the fixed 
point.\footnote{Actually, eq.~(\ref{Kmatrix}) need not be valid 
for $i,j,k$ with vanishing $Y^{ijk}$. However, 
the corresponding $a^{ijk}$ have homogeneous RG equations, and
do not contribute to the RG equations for gaugino masses, so they
decouple from this discussion.}
The solution to this equation will have the form
\beq
\pmatrix{ g_{\Agroup} M_{\Agroup} (Q) \cr a^{ijk}(Q) } =
\sum_n \left ( {Q_0 / Q}\right )^{\kappa_n} {\cal M}^{(n)}
\eeq
where $\kappa_n$ are eigenvalues of $K$ corresponding to eigenvectors
\beq
{\cal M}^{(n)} = \pmatrix{ {\cal M}^{(n)}_{\Agroup} \cr {\cal M}^{(n)ijk} }.
\eeq

In general the eigenvalues of $C$ do not correspond to
eigenvalues of $K$. However,
the traces of the matrices $C$ and $K$ 
do have the same real part.
Therefore, at least one of the eigenvalues $\kappa_n$ must have
a positive real part, since the sum of the eigenvalues of a matrix
is equal to its trace. This proves that the
running of the gaugino masses and the trilinear scalar couplings 
is again governed in the limit $Q \ll Q_0$ by a negative critical
exponent, corresponding to the eigenvector of $K $ with the eigenvalue
with the largest real part. This again implies power-law growth of all 
soft terms as one moves towards the IR within the fixed-point region. 

The behavior of soft  scalar squared mass RG evolution is more difficult
to ascertain. We can still obtain some insight, however, by considering
the form of the RG equations near the fixed point. We will begin by
considering this in the case of a general model with arbitrary gauge
and Yukawa couplings, and then specialize to the simpler case where Yukawa
couplings vanish (or can be safely neglected) at the fixed point. Even in
the case of a general model, the IR values of the subset of  soft squared
masses of scalars which do not have large Yukawa couplings at the
UV-stable fixed point will depend only on their gauge quantum numbers,
providing a possible solution to the supersymmetric flavor problem in the
MSSM. 

The scalar squared mass terms are
\beq
{\cal L} =  -(m^2)^j_i \phi^{*i} \phi_j \> .
\eeq
The RG equations for $(m^2)^j_i$ contain inhomogeneous terms quadratic
in gaugino masses and $a^{ijk}$ terms, and homogeneous terms.
Therefore, we can write
\bea
{d\over dt} (m^2)^{j}_i 
&=& 
\sum_{n,m}  \left ( {Q_0/ Q}\right )^{\kappa_n + \kappa_m^{*} }
[\Gamma_{n,m}]_i^j
-  L_{il}^{jk} (m^2)_k^l .
\label{bushisamaroon}
\eea
The fixed-point quantities  $[\Gamma_{n,m}]_i^j$ and
$L_{il}^{jk}$ can be given
exactly in
terms of first and second derivatives of the chiral superfield anomalous
dimension matrix $\gamma_i^j$ with respect to the couplings $g_{\Agroup}$, 
$Y^{ijk}$
and $Y_{ijk}$, using the results of \cite{Jack:1998iy}.

The general solution to eq.~(\ref{bushisamaroon}) takes the form
\bea
(m^2)^{j}_i 
&=& 
\sum_{n,m}  \left ( {Q_0/ Q}
\right )^{\kappa_n + \kappa_m^{*} }[X_{n,m}]_i^j
\nonumber \\   {} && 
+ \sum_N \left ( {Q_0/ Q} \right )^{\lambda_N} [X_N]_i^j ,
\label{gorehypocrite}
\eea
where $\lambda_N$ and $[X_N]_i^j$ are eigenvalues and eigenvectors 
of the matrix
$L_{il}^{jk}$, in the sense
\beq
L_{il}^{jk} [X_N]_k^l = \lambda_N [X_N]_i^j ,
\eeq
and $[X_{n,m}]_i^j$ are other constants fixed by the boundary conditions.

As one possible solution to the supersymmetric flavor problem,
suppose that a subset of the scalar fields including the squarks
and sleptons of the MSSM do not 
have large Yukawa couplings participating in the fixed point. 
The contributions to the RG eq.~(\ref{bushisamaroon})
for the
corresponding scalar squared masses
are then clearly flavor-independent. In the idealized limit
of a long running near the fixed point, 
$Q \ll Q_0$, one can write solutions in
terms of $\kappa$ (the largest of the ${\rm Re}[\kappa_n]$)
and $\lambda$ (the largest of the ${\rm Re}[\lambda_N]$).
For the flavor diagonal scalar squared masses
they will be of the form:
\bea
m^2_i(Q) &=& x_i \overline M_0^2 \left ({Q_0 / Q}\right )^{2\kappa}
\nonumber \\   {} && 
+ (y_i \overline m_0^2 - x_i \overline M_0^2)
\left ({Q_0 / Q}\right )^{\lambda}
\> .
\label{dimpledchad}
\eea
Here $x_i$ and $y_i$ are constants related to the eigenvectors of
the matrices $K$, $[\Gamma_{n,m}]_i^j$ and $L_{il}^{jk}$, 
while $\overline M_0^2$ and
$\overline m_0^2$ denote the overall scale of the boundary conditions for
holomorphic and non-holomorphic supersymmetry breaking terms,
respectively, at the scale $Q_0$.

By inspection of eq.~(\ref{dimpledchad}), 
there are two ways to have positive and 
flavor-conserving soft scalar squared masses in the
infrared:
\begin{itemize}
\item $2 \kappa > \lambda$ and $x_i M_0^2 > 0$,~~{ or}
\item $\lambda > 2 \kappa$ and 
$y_i \overline m_0^2 - x_i \overline M_0^2 > 0$.
\end{itemize}
Since the relevant quantities depend on the second derivatives
of anomalous dimensions at the fixed point, we do not see any 
reliable way to estimate them, or even determine their signs, 
far from the realm of perturbation
theory, in a completely
general theory. However, we do know that at least one ${\rm Re}[\kappa_n]$ 
is positive near
a UV-stable fixed point. 

If we further specialize to the case
that there is only one large gauge coupling and
no large Yukawa couplings at all at the fixed point,
then the anomalous dimension matrix and derivatives of it
are simultaneously diagonal, and
one has from \cite{Jack:1998iy} the exact formula:
\beq
L_{il}^{jk} = {{g^3/8\pi^2}\over (1 - g^2 C_G/8 \pi^2)} {\partial
\gamma_i \over \partial g} {C(k) \over d_G} \delta_i^j \delta_l^k ,
\eeq
where $d_G$ is the dimension of the adjoint representation.
It follows that all of the $\lambda_N$ vanish except for a single eigenvalue,
which can be evaluated as
\beq
\lambda = -{\partial \beta/\partial g} = \kappa.
\eeq 
This implies that the first of the above two
scenarios is the appropriate one, with the soft terms at low scale
being dominated by the gaugino mass contributions which
overwhelm any flavor-violating non-holomorphic contributions. One also
finds for the coefficient of $(Q_0/Q)^{2 \kappa}$ in the
first term in eq.~(\ref{bushisamaroon}) the exact result
\beq
\Gamma_i^j = {g\over 2} |M_0|^2 \left [
g {\partial^2 \gamma_i \over \partial g^2} +
\Bigl ({3 - g^2 C_G/8 \pi^2 \over 1- g^2 C_G/8 \pi^2}\Bigr ) 
{\partial \gamma_i
\over \partial g} \right ] \delta_i^j ,
\label{supergrass}
\eeq
where $M_0$ is the gaugino mass at $Q_0$.
In order to have a positive $m_i^2$ resulting in the IR from the
fixed-point running, it is necessary that the quantity in brackets
in eq.~(\ref{supergrass}) is negative for the corresponding scalar field
labelled by $i$. However, with present methods we
do not know how to evaluate the sign of this quantity even in the simplest
models, because it involves the second derivative of the anomalous
dimension.

Returning to the more general case with several relevant gauge groups
and/or Yukawa couplings, the critical behavior
may well still be dominated by gaugino contributions.
In practice, several eigenvalues $\kappa_n$  may have
real parts that are close together, so that the contributions
from several eigenvectors need to be included. Also, there may be unknown
non-perturbative corrections due to naively non-renormalizable terms in
the K\"ahler potential 
which become relevant at the fixed point. In order to
solve the supersymmetric flavor problem, however, a sufficient condition
is merely that a critical exponent related to gaugino masses dominates the
evolution of scalar squared masses
near the fixed point for a large enough range of scales.
Because the running is power-law rather than logarithmic, subdominant
flavor-violating contributions can be overwhelmed.

\section{Constraint from the conformal dimension of the gaugino bilinear}
\label{sec:dimconstraint}

In any theory with conformal invariance (supersymmetric or not), unitarity
requires that the
scaling dimension of any gauge-invariant scalar operator
$\phi$
must satisfy 
\beq
\Delta_\phi \geq 1,
\label{dnormal}
\eeq
with equality for a free field \cite{Mack}.
This gives strong contraints on possible fixed points. 
Before
asking how this impacts on the possibility of a UV-stable fixed point
in supersymmetry, we should note a slight subtlety. If $\phi$ is the scalar
component of a chiral superfield $\Phi$, then it is not always true 
that the ordinary scaling dimension $\Delta_\phi$ is equal
to the superconformal scaling dimension of the superfield, $D_\Phi$.
For example, they will differ if $\Phi$ appears in the superpotential,
as can be checked in the case of perturbative fixed points deformed
by Yukawa couplings in the IR superconformal window of SUSY QCD.
$D_\Phi$ participates in the essentially 
tautological rule of eq.~(\ref{dconsuper}),
whereas $\Delta_\phi$ satisfies the non-trivial rule of eq.~(\ref{dnormal}).

Typically, eq.~(\ref{dnormal}) will not yield a constraint for operators
$\phi$ built out of chiral fermions and scalars at a UV-stable fixed point,
since the anomalous dimensions are positive. However, in the case of
a simple gauge group with no superpotential,
the anomalous dimension of
the gaugino bilinear field $\lambda\lambda$ is given by 
\beq
\gamma_{\lambda\lambda} = \beta_M/M = -\kappa ,
\eeq
where the last equality only holds in the vicinity of the fixed point
and follows from the previous section.
Therefore the scaling dimension of the field is
\beq
\Delta_{\lambda\lambda} = 3 -\kappa \> .
\eeq
Enforcing the constraint eq.~(\ref{dnormal}), one finds
\beq
\kappa \leq 2 \> .
\label{kappaconstraint}
\eeq
Thus there is a constraint on the slope of the $\beta$ function at the
fixed point, corresponding also to a restriction on the critical
exponent for the gaugino mass. [In the case of an IR-stable fixed point,
$\kappa \leq 0$, so eq.~(\ref{kappaconstraint}) is automatically
satisfied.] More generally, the same argument implies that all eigenvalues
of the matrix $K$ of eq.~(\ref{Kmatrix}) satisfy
\beq
{\rm Re} [\kappa_n] \leq 2 .
\eeq

Note that $\lambda\lambda$ is also the lowest component of the chiral
superfield whose $F$-term is the gauge field strength. The superconformal
dimension $D$ of that chiral superfield is $3$, and its $R$-charge is 2,
as its $F$ component appears in the Lagrangian. This is consistent since,
as we noted above, the ordinary conformal dimension $\Delta$ of the
composite scalar field $\lambda\lambda$ can be and in fact  is distinct
from the superconformal scaling dimension $D$ of the chiral superfield to
which it belongs.

We emphasize that in a realistic theory, the superconformal symmetry is
actually broken by soft gaugino masses and small Yukawa and gauge
couplings, but this breaking does not mean that the constraint
eq.~(\ref{kappaconstraint}) does not apply. We can always imagine taking
the realistic model and simply turning off the small
superconformal-breaking effects; the constraint on the slope of the
gauge-coupling $\beta$ function then applies to the supersymmetric
couplings in that theory, and hence at least approximately to the
realistic model. 

\section{Hints from perturbative calculations}
\label{sec:hints}

It is difficult to make a definitive statement about the existence of a UV
fixed point even in SUSY theories, since this would require knowledge of
the $\beta$ function at all loop orders in theories for which the one-loop
contribution is positive.
However, it is still worthwhile to examine the 
perturbative expansions of $\beta$ functions in theories which could
have a UV-stable fixed point, as far as they are known.
For simplicity, let us first consider a SUSY gauge theory
with vanishing superpotential and a simple gauge group $G$.
One can expand the gauge coupling $\beta$
function as
\bea
{dg\over dt} = \beta =  
g \sum_{n=1}^{\infty} b^{(n)} 
\left ({g^2 \over 16 \pi^2}\right )^n .
\label{goresamoron}
\eea
The $b^{(n)}$ coefficients are scheme-dependent for $n\geq 3$, and are
known \cite{fourloops} in the dimensional reduction (DRED) \cite{DRED} and 
Novikov-Shifman-Vainshtein-Zakharov
(NSVZ)~\cite{NSVZ} schemes to four-loop order. In the NSVZ
scheme~\cite{fourloops}:
\bea
b^{(1)}  & = &  S_0-3C_G ;
\label{bone}\\
b^{(2)} & = &  4S_1+2C_GS_0-6C_G^2 ;
\label{btwo}\\ 
b^{(3)} & = & 
4 C_G^2 S_0
+  20 C_G S_1 
-12 C_G^3 
- 8 S_2 
- 4 S_0 S_1 ;
\label{bthree}
\\
b^{(4)} &=&
32 S_3
+ 48 (\zeta(3)-1) S_1^2
-(8 + 48 \zeta(3)) C_G S_0 S_1
\nonumber \\   &&
+ 8 C_G^3 S_0
+ 76 C_G^2 S_1
- 40 C_G S_2 
\nonumber \\   & &
-24 C_G^4
+ 8 S_0 S_2
- 4 S_0^2 S_1 ,
\label{bfour}
\eea
where 
$S_n\equiv \sum_r I_r C_r^n$ and the sum $\sum_r$ is over all chiral
superfield irreducible representations $r$.
The normalizations are such that in $SU(\Nc)$ with $\Nf$ flavors of
quark and anti-quarks,
\bea
C_G &=& \Nc;\\
S_n &=& \Nf \left ( {\Nc^2 -1 \over 2 \Nc} \right )^n.
\eea
In the DRED scheme, $b^{(1)}$ and $b^{(2)}$ are the same, but
\bea
b^{(3)}_{\rm DRED} & = &  
10 C_G^2 S_0 -21 C_G^3
+ 26 C_G S_1  \nonumber \\
& & - 8 S_2 - 6 S_0 S_1 - C_G (S_0)^2 \, ;\\
b^{(4)}_{\rm DRED} &=&
32 S_3
+ 48 (\zeta(3)-1) S_1^2
-102 C_G^4 \nonumber \\ &&
+ 54 C_G^3 S_0
+ 188 C_G^2 S_1  
- 80 C_G S_2 \nonumber
\\ & & -({184\over 3} + 48 \zeta(3)) C_G S_0 S_1
+ {64 \over 3} S_0 S_2 \nonumber \\  &&
-{14 \over 3} C_G^2 S_0^2
+ {4 \over 3} S_0^2 S_1
- {2 \over 3} C_G S_0^3 .
\eea
{}From eqs.~(\ref{bone})-(\ref{bfour}) we see that if $S_0 > 3 C_G$, then
the one
and two loop contributions to the $\beta$ function are positive.  However,
$b^{(3)}$ and $b^{(4)}$ can easily be negative, giving hope
that the gauge coupling will reach a fixed point with $\beta = 0$.  In
fact, this is
quite often the case at both three- and
four-loop order in both schemes. 

\begin{figure}[bt]
\centerline{\epsfxsize=3.0truein \epsfbox{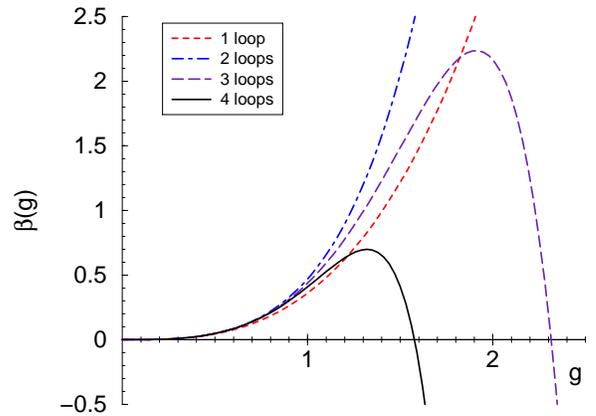}}
\vskip 0.0 cm
 \caption{The NSVZ gauge coupling $\beta$ function
of the example $SU(5)$ minimal missing partner model at 
1, 2, 3, and 4 loop orders.}
\label{betagraph}
\end{figure}

As a potentially realistic example, 
consider the minimal missing partner $SU(5)$ model \cite{missingpartner},
with $3 \times({\bf 5}
+ \overline{\bf 10})$, ${\bf 5} + \overline{\bf 5}$, $\bf{24}$, ${\bf
50}+\overline{\bf 50}$ and $\bf 75$ representations. The NSVZ $\beta$
functions at 1, 2, 3 and 4 loop orders are shown in
fig.~\ref{betagraph}. We see that including the three- and
four-loop terms, the $\beta$ function turns over and goes through zero.
Note that the terms in the $\beta$ function are not just alternating in sign;
both three- and four-loop contributions are negative.
The three- and four-loop UV fixed points are quickly reached in RG running
above the grand unified theory (GUT) scale $Q_{\rm GUT}$, so it is likely
that the true UV fixed point will be reached
(if it exists) significantly below the Planck scale.

Note that the negative slope of the $\beta$ function $\kappa =
-\partial\beta/\partial g$ obtained within the three- and four-loop
approximations is too large to satisfy the bound eq.~(\ref{kappaconstraint}).
Presumably, if this fixed point is indeed 
a feature of the full $\beta$ function,
the approach to the fixed point must be appropriately smoothed.

The DRED scheme $\beta$ functions yield qualitatively similar results,
although the locations of the fixed points are of course different.
Also, we
find that this behavior is quite generic in other GUT models with
large representations, including those based on $SO(10)$.
The same sort of thing happens as well in
models with several gauge group
factors and either large representations or large numbers of fundamental
representations. For example, in supersymmetric QCD with $\Nf$ flavors of
quarks and antiquarks, it is not hard to show that
the four-loop gauge $\beta$ function always has
a zero for $\Nf \geq 3 \Nc$. The result for three-loop beta functions 
depends on
the choice of NSVZ scheme or DRED scheme; in both cases a zero
of the $\beta$ function requires that the number of flavors $\Nf$
exceeds a critical value which can be obtained by solving a quadratic
equation. In the NSVZ scheme, the critical value for $\Nf$
is approximately $5\Nc$, while in the DRED scheme it is  close to $4 \Nc$.

It would be interesting (but of course hardly conclusive) to see whether
this behavior continues at higher loop orders.
Unfortunately, the calculation of $\beta$ functions at five or more loops
seems to be beyond present technology. The so-called ``exact"
$\beta$ functions of SUSY Yang-Mills theories actually rely on knowledge
of the anomalous dimensions of chiral superfields, which are only
available to three-loop order except in special cases. 
In any case, no finite order calculation of the $\beta$ function could prove
that a UV fixed point exists. Indeed, in the examples discussed here, the
perturbative expansion of the $\beta$ function shows no signs of
convergence in either the NSVZ or DRED schemes; the 1, 2, 3, and 4 loop
contributions are all numerically of roughly the same order near the
four-loop fixed point.  
However, the results above
illustrate that it is possible for the full $\beta$ function to rise as
$g$ does and then at some large value of $g$ to fall and cross zero,
thereby implying a UV fixed point. Note that this does not involve any
finetuning of parameters, but rather an assumption regarding the
qualitative behavior of the full $\beta(g)$.  The crucial question is
simply whether the exact $\beta$ function behaves qualitatively like the 1
and 2 loop approximations
in fig.~\ref{betagraph}, or like the 3 and 4 loop 
approximations.\footnote{There is also a logical possibility that
$\beta(g)$ remains non-zero for all $g$, but is small
enough that $g$ does not diverge at any finite energy scale. This would
correspond to a UV fixed point at infinite coupling. However, this seems
unlikely in four-dimensional SUSY models
because it would require the denominator of the NSVZ $\beta$ function
to be conspiratorially cancelled.}
We know of no existing argument or calculation
that definitively answers this question.

\section{The large-index limit}

The results of the previous section
suggest that we should probe the viability of  UV-stable
fixed points in the large-$\Nf$ limit, or more generally in the limit of
large Dynkin index for the chiral superfields, where some
all-orders results are now known. Bubble-sum calculations
have been used to find the anomalous dimensions and  gauge coupling
$\beta$ functions for SUSY QCD with a large number of flavors $\Nf$ in
both the DRED and NSVZ schemes, to all orders in the loop expansion and to
next-to-leading order in $1/\Nf$ \cite{FJJN}. These results 
can be written as a large-index limit in a general theory, in which only
terms in $b^{(n)}$ with $n-1$ factors of $S_i$ are kept for $n \geq 2$.
This calculation has a finite range of convergence $0 \leq g^2 < 24
\pi^2/S_0$ \cite{FJJN}. 
In this section, we examine whether these results can say
anything about UV-stable fixed points.

The results of \cite{FJJN} can be written in terms
of a rescaled\footnote{Note that our $g^2$ is equal to the $g^2/\Nf$
of Ref.~\cite{FJJN}, our $\hat K$ is the same, 
and our $S_0$ assumes the role of $\Nf$.}
coupling:
\bea
\hat K \equiv {S_0 g^2/ 16 \pi^2} 
\eea
and a special function 
\bea
G(x) = {\Gamma(2-2x) \over (\Gamma(1-x))^2 \Gamma(2-x) \Gamma(1+x)}.
\eea
Then, in a theory without Yukawa couplings, the anomalous dimension of
each chiral superfield is \cite{FJJN}
\bea
\gamma_r = - 2 C_r \hat K G(\hat K)/S_0 ,
\eea
to the leading order in the large-index expansion.
It follows that the NSVZ beta function is
\beq
\beta^{\rm NSVZ} = g \hat K \left [ {1 - 3 C_G/S_0 + 
4 (S_1/S_0^2) \hat K G(\hat K) \over 1 - 2 (C_G/S_0) \hat K }
\right ].
\eeq
The DRED beta function is given by a more complicated expression
\cite{FJJN}:
\bea
\beta^{\rm DRED} &=& g \hat K \Biggl [ 
1 - 3 C_G/S_0 + 2 (C_G/S_0) \int_0^{\hat K} G(x) dx \nonumber \\
&& + 4 (S_1/S_0) \int_0^{\hat K} (1-2x) G(x) dx \Biggr ] .
\eea
These results are exact to all-orders in $\hat K$, and to 
next-to-leading order in an expansion in $1/S_n$.
The interval of convergence for the infinite sum of bubble graphs is
$0 \leq \hat K < 3/2$, corresponding to a range over which $G(x)$
is finite.

In order to have a UV-stable fixed point, it is necessary to choose
$S_0 > 3 C_G$ and to find a zero of the beta function. In the NSVZ case,
this amounts to solving the equation
\beq
\hat K G(\hat K) = -{S_0^2 / 4 S_1} \> , 
\eeq
where we consistently drop contributions higher order in $C_G/S_0$.
This equation always has a solution for
some $\hat K$ between 1 and $3/2$, since $G(x)$ is continuous and
positive for $0 \leq x < 1$ and $G(1) = 0$ and $G(3/2) = 
-\infty$. This makes it appear that the large-index limit of the
theory always has a UV-stable fixed point! However, it can also be  
shown without much difficulty 
that $\beta^{\rm DRED}$ {\it cannot} have such a zero.

This apparent discrepancy between the  two schemes 
is due to the fact that the
putative UV-stable fixed point occurs in a regime in which the
next-next-leading order corrections (of order $C_G^2/S_0^2$, etc.)~are 
not negligible. 
Even though the sum to all orders in $\hat K$ is convergent, it
is not trustworthy in the relevant region. Furthermore, in specific models
one typically finds that the terms at 3 and 4 loop order which are the
most negative do not show up at all in the large index (large $\Nf$)
expansion. Therefore, we must conclude that this expansion
can make no unambiguous statement about the possibility of UV stable fixed
points.

\section{UV stable fixed points cannot occur with perturbative
couplings}

In the case of IR-stable fixed points, one can often tune the number of
chiral fields to obtain arbitrarily small critical couplings, so that
the existence of the fixed point can be reasonably established in some
neighborhood within the
space of models \cite{Banks:1982nn}. 
This is unfortunately not possible for UV-stable fixed
points in supersymmetric theories. First consider the case in which
there is no superpotential. In general, the existence of
a perturbative fixed point requires that the one- and two-loop
contributions in eq.~(\ref{goresamoron})
have opposite signs and $|b^{(1)}| \ll |b^{(2)}|$, 
so that 
\beq
{g^2_* \over 16 \pi^2} \approx -b^{(1)}/b^{(2)}  \ll 1 .
\label{gpert}
\eeq 
However, in supersymmetric
theories the
two-loop $g^5/(16 \pi^2)^2$ coefficient in the
beta function is strictly larger than the one-loop $g^3/16 \pi^2$
coefficient; $b^{(2)} = b^{(1)} + 4 S_1$. 
Therefore, if one tunes the one-loop coefficient to be either zero
or small and positive, the two-loop contribution will necessarily
be large and positive, since there is no way to tune $S_1$ to be small. 
This means that eq.~(\ref{gpert}) cannot be satisfied.
Furthermore, it means that 
the location of a possible UV-stable fixed point cannot
be made perturbative by balancing a large negative 3-loop contribution
against controllably small one- and two-loop contributions.

It might seem that this argument could be evaded by introducing
Yukawa couplings in a superpotential. Since Yukawa couplings enter
at two loops into the gauge coupling $\beta$ function, this naively appears
to allow the possibility of tuning both one- and two-loop to be small.
However, the $\beta$ functions for those Yukawa couplings
also must vanish at the fixed point.
We can prove quite generally that this cannot happen with
UV stability, as follows.

Consider a general superpotential of the form given in
eq.~(\ref{superpot}).
The $\beta$ function for the summed squares of all Yukawa couplings
is
\beq
{d \over dt} (Y^{ijk} Y_{ijk}) = 6 Y_{ijl} Y^{ijk} \gamma^l_k
\label{goreisapathologicalliar}
\eeq
where $\gamma^l_p$ is the anomalous dimension matrix, and the indices
$i,j,k,\ldots$ run over all chiral supermultiplet degrees of
freedom. At
one loop
order, one has 
\beq
16 \pi^2 \gamma^{(1)k}_l  = {1\over 2} Y_{ijl} Y^{ijk} - 2 \sum_{\Agroup} 
g_{\Agroup}^2
C_{\Agroup} (k) \delta^k_l ,
\eeq 
where the sum $\sum_{\Agroup}$ is over simple and $U(1)$ gauge
groups, and $C_{\Agroup}(k)$ is the quadratic Casimir invariant of the chiral
superfield carrying the index $l$. Requiring that the
$\beta$ functions of the gauge couplings $g_{\Agroup}$ vanish, and
rewriting eq.~(\ref{goreisapathologicalliar}) 
as a sum over the irreducible representations of
chiral superfields, chosen so as to diagonalize the matrix
$\gamma_l^{(1)p}$ at the fixed point, one obtains to leading order:
\beq
{d\over dt} (Y^{ijk} Y_{ijk}) = 12 \Bigl (16 \pi^2 \sum_r d_r|
\gamma^{(1)}_r|^2 
+ \sum_{\Agroup} d_{\Agroup} b_{\Agroup}^{(1)} g_{\Agroup}^2 \Bigr ),
\label{pregnantchad}
\eeq
where $d_{\Agroup}$ 
is the dimension of the adjoint representation of the subgroup
labelled by $\Agroup$, and the $\sum_r$ is over the irreducible
representations of the chiral superfields with representation dimension $d_r$.
At a fixed point, eq.~(\ref{pregnantchad}) must vanish.
This requires that some $b_{\Agroup}^{(1)} < 0$ in order for 
a cancellation to occur on
the right-hand side.
So a perturbatively-reliable fixed
point is only possible if a gauge-coupling $\beta$ function is 
significantly negative near
weak coupling, implying
an ultraviolet free theory or an unstable fixed point in the UV.
We therefore conclude that stable UV fixed points cannot
occur with perturbative couplings.

\section{Constraints from positivity of central charges}
\label{sec:central}

In this section we consider constraints following from
the positivity of coefficients in the stress-energy trace anomaly.
In an external supergravity background with sources
for conserved flavor currents, the trace anomaly contains terms
proportional to the square of the dual of the Riemann
curvature, the square of the Weyl tensor,  
and the square of the flavor symmetry field strength.
At a superconformal fixed point, the central charge coefficients
for these terms can be shown to be respectively \cite{Anselmi:1998ys}:
\bea
a &=& {3 \over 32} \left ( 2 d_G + {\rm Tr} [(1-R_i) (1 - 3 (1-R_i)^2)]
\right ) \\
c &=& {1 \over 32} \left ( 4 d_G + {\rm Tr}[ (1-R_i) (5 - 9 (1-R_i)^2)]
\right ) \\
b &=& 3{\rm Tr}[ (1 - R_i) F^2_i] .
\eea
Here the sums are over all chiral superfields with
$R$-charges associated with the superconformal algebra
$R_i$, flavor symmetry
charges $F_i$, and $d_G$ is the dimension of the adjoint
representation of the gauge group.
At a non-interacting (free) fixed point, all of the $R_i$ should
be replaced by
$2/3$, and at an interacting fixed point they are constrained by the
fact that the superconformal scaling dimension is $D = 3R/2$ for 
gauge-invariant chiral
superfields.

At a superconformal fixed point, each of the coefficients $a$, $b$, and
$c$ must be  positive \cite{Anselmi:1998rd,Anselmi:1998ys}. 
In addition, 
Cardy's conjecture
\cite{Cardy} implies that
the generalization of Zamolodchikov's $c$-theorem in 2 dimensions
should be that the coefficient $a$ is larger in the
UV than in the IR. This expresses the fact that $a$ is a 
suitably-weighted measure of the
number of degrees of freedom, which are irreversibly integrated out in the
Wilsonian approach. In our case, the UV fixed is supposed to be 
interacting,
and the IR fixed point is free. The constraints $a_{\rm IR}>0$, 
$b_{\rm IR}>0$
and $c_{\rm IR} > 0$ are therefore automatically satisfied. There 
are, however,
non-trivial constraints following from $b_{\rm UV} > 0$ and 
$c_{\rm UV} > 0$,
and, assuming Cardy's conjecture is true,
 $\Delta a \equiv a_{\rm UV} - a_{\rm IR} > 0$. 
The last condition is equivalent to
\beq
\Delta a = {1\over 96} {\rm Tr}[ (3 R_i-2)^2 (3 R_i - 5)] > 0 .
\label{cardya}
\eeq
Note that the left side of this inequality has the opposite sign from that
used in Ref.~\cite{Anselmi:1998ys}, since the correspondence between 
(UV, IR) and (free, interacting) fixed points has been switched.

In order to satisfy eq.~(\ref{cardya}), clearly
at least one chiral superfield should have $R$-charge greater than $5/3$.
This constraint, if it should really be imposed, is quite strong and
would readily rule out UV-stable fixed points in many models. For
example, consider SUSYQCD with $\Nf \geq 3 \Nc$ flavors.
In that case, anomaly cancellation requires that $R_Q = R_{\overline Q} = 
1 - \Nc/\Nf$. It follows that
$a$, $b$, and $c$ are each positive. However, since $R_Q$ 
is not as large as 5/3, eq.~(\ref{cardya})
cannot be satisfied. Thus the rigorous exclusion or establishment of the
existence of
a UV-stable fixed point in this model would be a non-trivial test
of Cardy's conjecture.

Let us assume that Cardy's conjecture is true, and explore under what
circumstances a more complicated theory could satisfy it at a UV-stable
fixed point.
Even if a theory can be arranged to have a non-anomalous $R$-symmetry
with one or more chiral superfields with $R_i > 5/3$, the rigorous
constraint $b_{\rm UV} > 0$ is then quite strong. For example, suppose that
the theory includes chiral superfields $\Phi_1$ and $\Phi_2$
with the same gauge quantum numbers and $R_1, R_2 > 5/3$. Then there is
a non-anomalous $U(1)$ flavor symmetry under which $\Phi_1$ and $\Phi_2$
have charges $F_1=1$ and $F_2=-1$ respectively 
and all other chiral superfields
are neutral. This clearly implies the inconsistency $b_{\rm UV} < 0$ for
this flavor symmetry.
 
The simplest model that apparently can satisfy all constraints including
eq.~(\ref{cardya}) is
supersymmetric QCD
with gauge group $SU(\Nc)$, $\Nf$ flavors of quarks and
antiquarks $Q+\overline Q$ in
the
fundamental + anti-fundamental representations, an adjoint $A$, and
a singlet $S$. The anomaly cancellation condition for the $R$-charges
is
\beq
R_Q = R_{\overline Q} = 1 - {\Nc\over \Nf} R_A .
\label{foofighters} 
\eeq
Eq.~(\ref{cardya}) can only be satisfied if
$R_A > 5/3$. Now, if there were no
superpotential, then there would be an anomaly-free $U(1)_F$ symmetry
under
which
$(Q, \overline Q, A)$ have charges $(1,1, -\Nf/\Nc)$. The corresponding flavor
symmetry central charge constraint $b_{\rm UV} > 0$ would amount to
\beq
2 \Nc \Nf (1 - R_Q) + (1 -{1\over \Nc^2}) \Nf^2 (1-R_A) > 0 ,
\eeq
and this cannot be made consistent with
eqs.~(\ref{cardya}) and (\ref{foofighters}). 
However, if we introduce a superpotential
of the form
\beq
W = S Q \overline Q + S^3 \, ,
\label{supersave}
\eeq
then $U(1)_F$ is removed and there is no $b_{\rm UV}$ constraint.
The superpotential eq.~({\ref{supersave}) also fixes the $R$-charges to 
have the unique values
\bea
R_S &=& R_Q = R_{\overline Q} =2/3 \, ; \\
R_A &=& {\Nf \over 3 \Nc}.
\eea
Therefore one must have $\Nf > 5 \Nc$ in order to have a UV-stable
fixed point consistent with Cardy's conjecture. 
Writing $\Nf/\Nc = 5 + p$ with positive $p$, one finds
\bea
\Delta a &=& (\Nc^2-1) {(3+p )^2 p\over 96} > 0 \, ; \\
c_{\rm UV} &=& {5+p\over 96}  \left [
(\Nc^2-1) (2 + p +p^2) + 8 \Nc^2
\right  ] > 0 ,
\eea
so all constraints are satisfied. The scaling dimensions of
all gauge-invariant chiral superfields are at least 1 in this model. For
example, the
$R$-charges of gauge invariant
chiral superfields are $R_{Q\overline Q} = 4/3$, $R_S= 2/3$, $R_{A^2} = 
2 \Nf/3\Nc$, leading to scaling dimensions 2, 1, and $\Nf/\Nc > 5$
respectively. 

A similar model consists of $SU(\Nc)$ with $\Nf$ flavors of
quarks and antiquarks $Q,\overline Q$ 
and two adjoints $A_1$ and $A_2$. The
superpotential at the putative fixed point is taken to be of the
form
\beq
W = A_1 Q \overline Q + A_1^3 \> .
\eeq
This fixes the anomaly-free $R$-charges to be
\bea
R_{A_1} &=& R_Q = R_{\overline Q} =2/3 ;\\
R_{A_2} &=& {\Nf +\Nc\over 3 \Nc}.
\eea         
All of the constraints are satisfied for $\Nf > 4 \Nc$ in this model.

As these examples illustrate, it is non-trivial to satisfy both
$\Delta a > 0$ and $b_{\rm UV} > 0$ at a UV-stable fixed point. One must
generally be attracted to a superpotential that eliminates dangerous 
flavor symmetries that could otherwise 
enter into the $b_{\rm UV} > 0$ constraint,
while allowing or enforcing $R_i > 5/3$. Evidently, the RG flow into the
UV will either not approach a fixed point, or will be attracted towards
a special superpotential with the right properties. The superpotential
that the theory is attracted towards in the UV may bear little resemblance to 
the IR superpotential.

Many more complicated examples can be constructed. For example,
returning to the minimal missing partner $SU(5)$ model examined in section
\ref{sec:hints}, one finds that $\Delta a > 0$ and $b_{\rm UV} > 0$
for all flavor symmetries cannot simultaneously 
be achieved unless a superpotential is introduced. One way to
proceed is to choose a superpotential that involves every chiral
superfield except the adjoint, which then obtains an $R$-charge greater
than $5/3$. In models with many higher-dimensional representations,
it is often possible to find a superpotential which is
a candidate for attraction to a UV-stable fixed point.

It should be emphasized that Cardy's constraint
$\Delta a > 0$ remains as a conjecture rather than a rigorously proved
theorem. There are now a large number of examples
\cite{Cardy,Cardyexamples,Anselmi:1998ys} in
which it is satisfied, and no other candidate for the
4-dimensional version of the $c$-theorem in 2 dimensions appears to
acceptable.  However, the known examples in which it has been checked are
all free theories in the UV,  and it
is a possibility that it could be violated in theories with
UV-stable fixed points. Therefore it remains an open question whether the
constraint $\Delta a> 0$ should necessarily be imposed. It seems an
important challenge to see whether this can be made rigorous.

\section{Concluding remarks}

We have investigated the hypothesis that a non-asymptotically free
supersymmetric field theory flows to a fixed point in the ultraviolet.
At one- and two-loop order, the gauge $\beta$ functions do not hint at
this possibility, but fixed point indications do appear at
three- and four-loop level.  Although no perturbative calculation
can prove the existence of a UV fixed point, the qualitative indications
of higher-order calculations have inspired us to look more carefully at
the possibility.

Starting with the fixed point assumption, the many exact 
results~\cite{Shifman:1997ua,Arkani-Hamed:2000mj} of 
supersymmetric field theories 
enable us to investigate the consistency and implications of the claim.
The most important rigorous consistency check is the requirement that all
gauge-invariant and Lorentz invariant operators must have scaling dimension
greater than or equal to 1.  This places a constraint on the slope of the
gauge beta function, and ultimately on the strength of the power-law running
for soft supersymmetry breaking parameters.  The other important constraints
arise from the positivity of trace anomaly coefficients, including
the Cardy conjecture.  These constraints, if indeed they are rigorously
required, rule out some otherwise-consistent proposals for 
UV fixed point theories
(e.g., simple SUSY QCD with $\Nf > 3 \Nc$); 
however, they do not rule out the existence of
UV fixed point theories in general, as we demonstrated by exhibiting 
viable examples.

The most important implication of a fixed point theory is the power-law
running of the soft supersymmetry breaking masses.  Going up in energy,
the soft masses power-law run to small values in the UV fixed point
region.  For some applications, it is more convenient to think in the
other direction.  Namely, extremely small supersymmetry breaking
parameters in the far UV power-law run to large (weak scale) values in the
IR.  This flow could generate a large hierarchy between the original
supersymmetry breaking masses and the weak scale. If the flow is dominated
by gaugino masses, the scalar masses will be flavor-blind due to the
running near the superconformal UV-stable fixed point, yielding a solution
to the supersymmetric flavor problem. (In some sense, this is an opposite
scenario to that of \cite{nelsonstrassler}, which invokes an 
infrared-stable fixed point to solve the SUSY flavor problem.) 

We
note that near the fixed point, non-renormalizable operators may become
relevant due to large anomalous dimensions. For superpotential operators,
this seems unlikely since the anomalous dimensions are typically positive.
However, we have no such control over K\"ahler potential terms. Therefore, a
true solution to the supersymmetric flavor problem requires the (not
unreasonable, we believe) assumption that these are still subdominant 
compared to contributions from power-law running of flavor-blind gaugino
masses.

Such a large hierarchy may also have implications for the gravitino
problem~\cite{gravitino}, since the ultimate supersymmetry-breaking
order parameter $F$ could actually be much smaller than the
power-law-enhanced masses of the MSSM superpartners. Also, if the
gravitino is light enough, superpartners could decay into it
promptly, leading to collider signatures characteristic of gauge
mediation~\cite{gaugemediation}: e.g.~hard photons  plus missing energy,
or hard leptons plus missing energy. Because of the constraint on the
anomalous dimension of gaugino masses discussed in section IV, this would
require that the theory remains near the UV-stable fixed point over a range
of RG scales corresponding to $\Delta t$ significantly larger than 
${\rm ln}( M_{\rm GUT}/M_{\rm Planck})$.

It may be difficult to uniquely identify a superpartner mass spectrum as
arising from an ultraviolet fixed point. The $SU(5)$ GUT missing partner
model example that we briefly considered in the text might look
qualitatively like any other theory with unified gaugino masses. However,
it is also quite possible that the power-law running in the UV at very high
scales would generate a different scalar superpartner spectrum than the
ones from gauge mediation or the usual assumptions of minimal
supergravity, and the ideas could in principle be resolved. To do this in
detail would require an improved understanding of the non-perturbative
running of soft scalar masses near the fixed point.

\vspace{0.2in}

\noindent
{\it Acknowledgments: }
We are grateful to N.~Arkani-Hamed, W.~Bardeen and M.~Shifman
for helpful conversations.
This work was supported in part by National Science Foundation Grant
No.~PHY-9970691 and by the Alfred P. Sloan Foundation.

\def\Journal#1#2#3#4{{#1} {\bf #2}, #3 (#4)}
\def\add#1#2#3{{\bf #1}, #2 (#3)}

\def\NPB{{\em Nucl. Phys.} B}
\def\PLB{{\em Phys. Lett.}  B}
\def\PRL{{\em Phys. Rev. Lett.}}
\def\PRD{{\em Phys. Rev.} D}
\def\PR{{\em Phys. Rev.}}
\def\ZPC{{\em Z. Phys.} C}
\def\SJNP{{\em Sov. J. Nucl. Phys.}}
\def\AnnP{{\em Ann. Phys.}}
\def\JETPL{{\em JETP Lett.}}
\def\LMP{{\em Lett. Math. Phys.}}
\def\CMP{{\em Comm. Math. Phys.}}
\def\PTP{{\em Prog. Theor. Phys.}}

\end{document}